\documentclass[onecolumn,showpacs,amsmath,amssymb,preprint,tightenlines,10pt]{revtex4}
\pdfoutput=1
\usepackage{graphicx,anysize}
\usepackage{bm,bbm,epstopdf}
\usepackage{multirow}
\usepackage{amsmath}
\usepackage{bbm}
\usepackage{amssymb}
\usepackage{threeparttable}
\usepackage{float}
\usepackage{varioref}
\usepackage{epic}
\usepackage{eepic}
\usepackage{makeidx}
\usepackage{epsfig}
\usepackage{subfig,url}
\usepackage{supertabular}

\begin{document}
\preprint{DOI: 10.1039/c2sm07384f, Soft Matter Communications}
\bibliographystyle{prsty}

\title{Role of disclinations in determining the morphology of deformable fluid interfaces}

\author{N. Ramakrishnan }
\email{ram@physics.iitm.ac.in}
\affiliation{Department of Physics, Indian Institute of Technology Madras, Chennai  600036, India}
\author{John H. Ipsen}
\email{ipsen@memphys.sdu.dk}
\affiliation{MEMPHYS- Center for Biomembrane Physics, Department of Physics and Chemistry, \\
University of Southern Denmark, Campusvej 55, DK-5230 Odense M, Denmark } 
\author{P. B. Sunil Kumar}
\email{sunil@physics.iitm.ac.in}
\affiliation{Department of Physics, Indian Institute of Technology Madras, Chennai  600036, India}

\date{\today}
\begin{abstract}
 We study the  equilibrium shapes  of  vesicles, with an in-plane nematic order, using a Monte-Carlo scheme  and show that  highly curved shapes, like tubes and discs, with a striking similarity to the structures engendered by certain curvature sensing peripheral  membrane proteins,  can be spontaneously generated by anisotropic directional curvature with nematic disclinations playing and important role.  We show that the coupling between nematic order and local curvature could lead to  like defects moving towards each other and  unlike defects moving away, in turn leading to tube formation.    Thermally induced defect pair production  lead to branched tubular structures.  It is also shown that helical arrangement of the membrane tubes, with nematic field spiraling around it, is a dominant soft mode of  the system.
\end{abstract}

\pacs{{PACS-87.16.D-,} {Membranes, bilayers and vesicles.} 
                           {PACS-05.40.-a} {Fluctuation phenomena, random processes, noise and Brownian motion.} 
  	                   {PACS-05.70.Np} {Interfaces and surface thermodynamics}    }

\keywords{vesicles -- membranes -- bending elasticity -- statistical mechanics --  --  membrane shape -- conformational
      fluctuations -- Monte-Carlo integration -- liquid crystals  --  -- XY model -- Lebwohl-Lasher model}

\maketitle

 Fluid interfaces with an inplane orientational order can sustain  amazingly complex morphologies.  Apart from the  interesting physics  that  it  can offer, in terms of disclination dynamics on deformable surfaces,  study of such interfaces  are also contribute to understanding morphologies of cellular organelles.
 
The organelles of a biological cell  have membranes with highly curved edges and tubes, as  seen in the Endoplasmic reticulum, the Golgi  and the inner membrane of mitochondria. Tubulation has also been observed, {\it in vitro}, in self assembled systems of pure lipids \cite{Markowitz:1991p3276}. It has been shown  that, macromolecules,  which  constitutes and decorates the membrane surface, strongly influence the morphology of membranes. For instance  proteins from the dynamin superfamily  are known to pull out membrane tubes  while  oligomerizing themselves  into a helical coat along the tube\cite{Praefcke:2004p3309}. The BAR domain containing proteins in general can  induce a wide spectrum of membrane shapes ranging from protrusions to invaginations depending on the geometry and interaction strength of the BAR domain \cite{Zimmerberg:2006p510,Voeltz:2007p1399,Shibata:2009p643}.

The standard Helfrich model\cite{Helfrich:1973p693} for membranes, based on mean curvature energy, cannot explain the stability of such highly curved structures. Existence of anisotropic bending energy will be the minimal requirement to explain  the stability of tubular shapes, which could arise from an in-plane orientational field on the membrane \cite{Fournier:1996p488,Fournier:1998p2999}.   Planar orientational order could be intrinsic to the membrane~\cite{KraljIglic:2006p3974}, due to the structural properties of its constituents, or could arise as a result of membrane interactions with external agents~\cite{Zimmerberg:2006p510,Voeltz:2007p1399,Shibata:2009p643}. Lipid molecules  tilted with respect to the layer normal, as seen in the $L_{\beta}^{'}$~\cite{Smith:1988p3535} and $P_{\beta}^{'}$~\cite{Tardieu:1973p711} phases, is a well known case of intrinsic  in-plane orientational order.   Anisotropic  peripheral proteins ({\em curvactants}),  inducing  membrane deformations \cite{Yin:2009p255}  either by scaffolding or by insertion of their amphipathic helices into the membrane~\cite{Zimmerberg:2006p510}, are examples for extrinsic sources of orientational order.  In general the order can be represented by a {\em $p$-atic} surface field, invariant under rotations by $2\pi/p$. Equilibrium shapes of a surface with  {\em $p$-atic} in plane order, have been investigated earlier~\cite{Park:1992p946,MacKintosh:1991p685}.  Mean field phenomenological models of orientation order on membrane surface, in the context of lipid tilt and chirality, have predicted the stability of tubular membranes and helical ribbons\cite{Helfrich:1988p3004,Nelson:1992p9,Selinger:1993p108,Schnur:1993p1277}. More recently, a mesoscopic particle model was used to show that presence of BAR domains on membrane surfaces can  lead to  tubulation and vesiculation~\cite{Ayton:2007p3485,Ayton:2009p3487}.

The coupling of curvature to orientational order leads to interesting defect dynamics and  thermally excited shapes. Thus in addition to  its biological relevance, membranes with in plane order also  provides a unique medium to study disclination  dynamics on deformable surfaces.  So far such studies have been carried  out only on surfaces with prescribed curvature~\cite{Shin:2008p174}.  Apart from the interesting physics that it can generate these investigations  may also provide new routes to new functional materials\cite{Nelson:2002p335}. 

 In this paper we use the Monte Carlo model developed by us~\cite{Ramakrishnan:2010p1249, supp} to explore the spectrum of vesicle shapes engendered  by an in-plane nematic field ${\hat m}$.  The technique developed by us enables one to go beyond the mean field regime and also to explore the full nonlinear regime.  We scale all energies by $k_BT$.  A moderate mean curvature stiffness  $\kappa=10$ is chosen to  enable study of thermally induced shape fluctuations. Higher values of $\kappa$ are chosen to study equilibrium shapes in the absence of  thermal fluctuations.  We choose the directional curvature moduli, along and perpendicular to the orientational field, as  $\kappa_{\parallel}=5$ and  $\kappa_{\perp}=0$ respectively.  The results with $\kappa_{\perp}\ne 0$ is qualitatively same and is not discussed separately.   We vary  the  spontaneous curvature  along the in-plane field $C_{0}^{\parallel}$, to explore different shapes. 
The in-plane  nematic field, distributed uniformly over the entire surface, is constrained to be in the ordered phase by setting the Lebwohl Lasher coupling constant  $\epsilon_{LL}=3k_{B}T$\cite{Lebwohl:1972p426}.  For this  values of the coupling constant,  with  $\kappa_{\parallel}=0$, the  nematic field orients itself on a tetrahedron with four $+1/2$ disclinations, as predicted in \cite{Park:1992p946}.  We will first investigate the effect of the directional spontaneous curvature on the equilibrium membrane conformations when the $\kappa$ value is high so that thermal undulations can be neglected.

\begin{figure}[!h]
\centering
\includegraphics[width=5in]{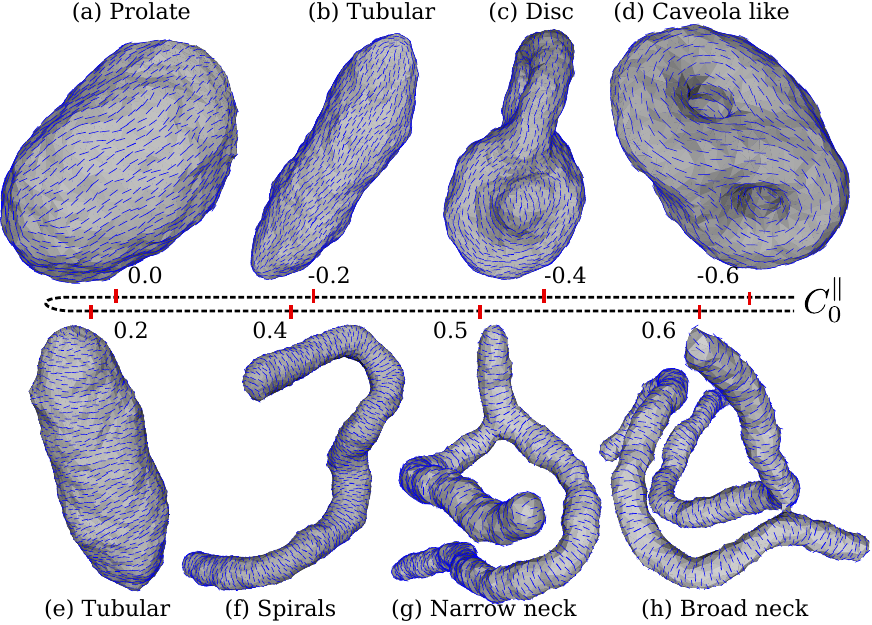}
\caption{\label{fig:kperp0} Membrane conformations for different values of  $C_{0}^{\parallel}$ with $\kappa=10$, $\kappa_{\parallel}=5$, $\kappa_{\perp}=0$ and $\epsilon_{LL}=3$. ( See supplementary materials for high resolution figures of nematic field arrangement.)}
\end{figure} 
At high $\kappa$ thermal undulation does not change the shapes significantly and the shapes obtained from the equilibrium simulations will be  the same as that from a mean field calculation.  It is known that  dynamic triangulation Monte Carlo (DTMC) models,  based on Helfrich Hamiltonian~\cite{Helfrich:1973p693}, do reproduce the mean field phase diagram of vesicles~\cite{Gompper:1994p205}.  Here we  determine  the equilibrium shapes of   {\it decorated} vesicles with an in-plane orientational  order, also taking thermal fluctuations into account .   As can be seen from Fig.\ref{fig:kperp0},  we see tubular and disc like shapes emerging,  even when  there  are no constraints on the  area to volume ratio of the vesicle.   Directional spontaneous curvature and interaction between the disclinations  are two important factors that  determine the shape of these decorated vesicles.  Since the membrane is self  avoiding the  constraint imposed by the fixed  topology  also plays a role.  
\begin{figure}[!h]
\centering
\includegraphics[width=2.75in]{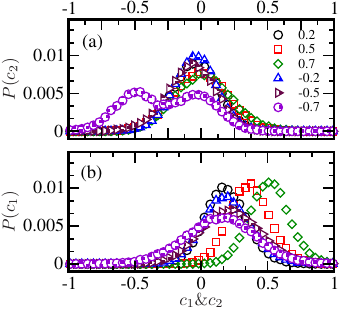}
\caption{\label{fig:princurvdist}(Color online) (a,b) Distribution of the maximum and minimum principal curvatures, $c_{1}$ and $c_{2}$ respectively, for different values of $C_0^{\parallel}$. }
\end{figure}

\begin{figure}[!h]
\centering
\includegraphics[width=3.0in]{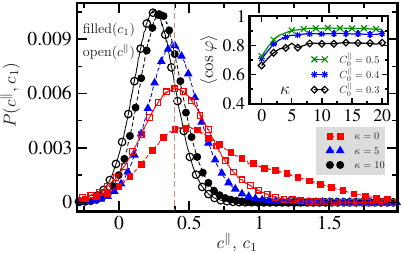}
\caption{\label{fig:pcur-dcur}(Color online) Comparison of the $c_1$   (filled symbols) and directional curvature  $c^{\parallel}$ (open symbols) distribution, at $C_{0}^{\parallel}=0.4$, $\kappa_{\perp}=0$, for different values of $\kappa$.  Inset shows the change in  the angle $\varphi$ as a function of $\kappa$.  Dotted vertical line marks  $c_1=C_{0}^{\parallel}=0.4$.}\end{figure}

Four curvatures can be identified, the curvatures along the principal directions  $c_1, c_2$ with $c_1>c_2$ and the directional curvatures, $c^{\parallel}, c^{\perp}$, along and perpendicular to ${\hat m}$.    Fig.\ref{fig:princurvdist} shows the distribution $P(c_i)$, of $c_{i=1,2}$, on a vesicle,  for different $C_0^{\parallel}$ values when $\kappa_{\perp}=0$.  It is clear from  Fig.\ref{fig:princurvdist}(a \& b) that for $C_0^{\parallel}>0$  the distributions  have  a single  peak, with  $P(c_1)$, peaked close to, but less than,  the value of  $C_0^{\parallel}$, while $P(c_2)$  is peaked at zero. The resultant  value of the principal curvature  arising from  the competition between the resistance to bending imposed by $\kappa$ and the directional spontaneous curvature set by $C_0^{\parallel}$, is  shown in Fig.\ref{fig:pcur-dcur}.   When  $\kappa$ decreases  the peak position    of $P(c_1)$  moves toward that of $C_0^{\parallel}$ and then to a value higher than  $C_0^{\parallel}$  for $\kappa \sim 0$. Note that for $\kappa=0$  there are no competing elastic forces and the desired  directional curvature  can be achieved by setting ${\hat m}$ at an angle $\varphi\ne0$ with respect to the direction of $c_1$. For  a tube, a non zero value of $\varphi$  implies a configuration with   nematic spiraling around the tube.  The angle $\varphi$ is shown as inset in Fig. \ref{fig:pcur-dcur}. As expected this angle is a decreasing function of the bending modulus $\kappa$  and saturates to zero for large bending rigidity.

Negative spontaneous directional curvature ($C_0^{\parallel}<0$) induces tubes to form into the vesicle.  However, restrictions  imposed by the membrane self avoidance and topology prevent a complete inversion of the membrane through inward tubulation.  This leads to a bimodal  distribution of $c_2$, with a peak at $c_2=0$ in addition to  the expected peak at $c_2 \approx C_0^{\parallel}$ (see  Fig.\ref{fig:princurvdist}~a).  This  bimodal peak arises due to presence of inward tubes (with most of the vertices with $c_2<0, c_1=0$ ) and an outward curved  surface (with most vertices having $c_1>0, c_2=0$).

Another important factor that determines the shape of a  vesicle is the positioning of disclinations.  We know that, for a nematic field on  a closed surface with spherical topology,    the total disclination strength  should be  2.   There is no such topological restriction on the number of defects themselves and thermal fluctuation could excite oppositely charged defect pairs.  

\begin{figure}[!h]
\centering
\includegraphics[width=3.5in]{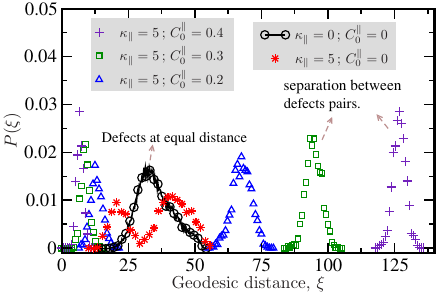}
\caption{\label{fig:geodesplot}(Color online) Distribution of the nematic defect separation  ($\xi$) for $\kappa=10$, $\epsilon_{LL}=3$ and $\kappa_{\perp}=0$.}
\end{figure}

\begin{figure}
\centering
\includegraphics[width=5in]{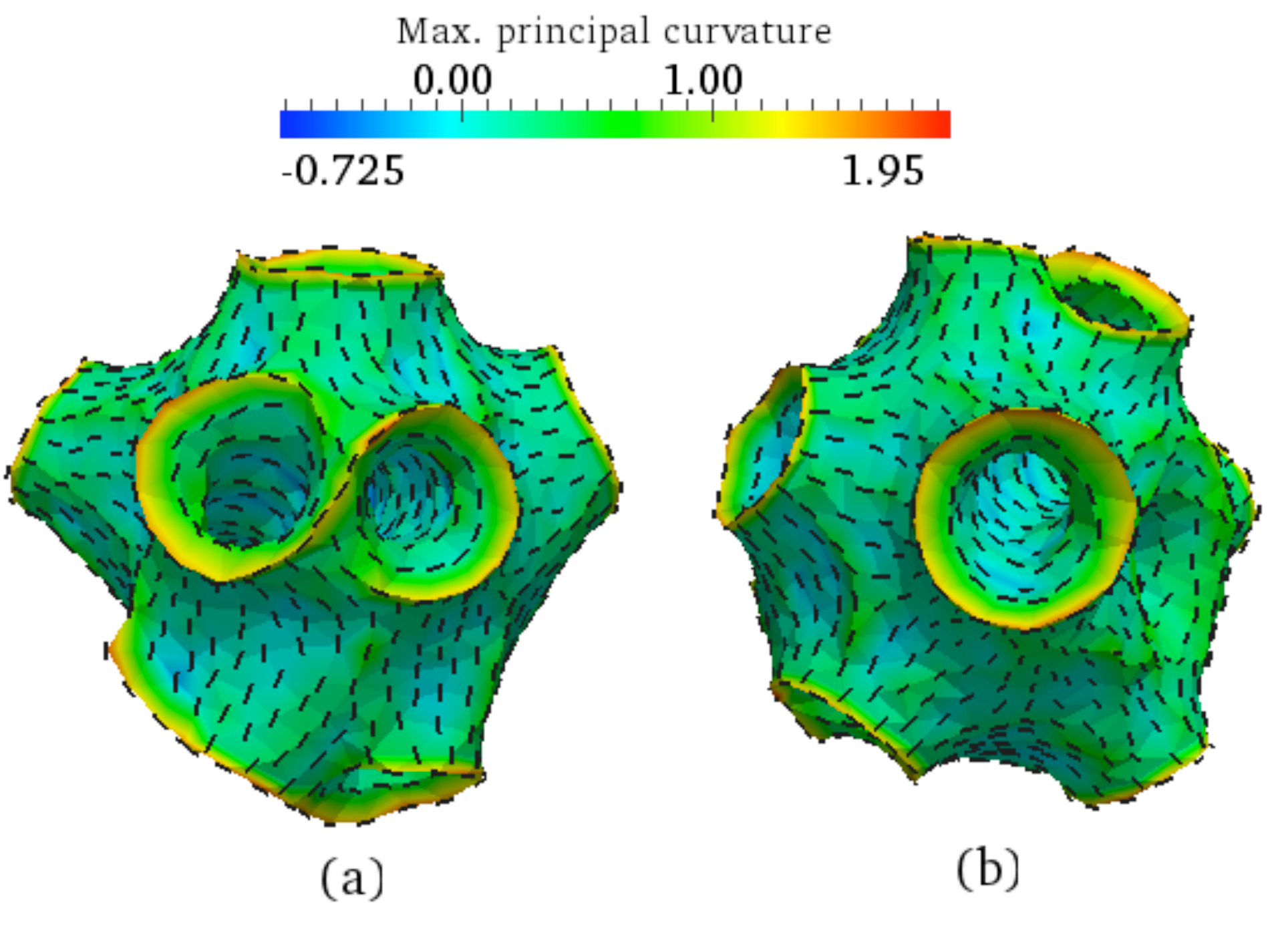}
\caption{\label{fig:line} (Color online) Line discontinuities in the nematic orientation for $C_0^{\parallel}<0$. (a) Paired tubes have rims that end on $-1/2$ defects while (b)  closed rims are surrounded by four $-1/2$ defects (b).}
\end{figure}

In the absence of any stiffness-orientation coupling,  the energy of a pair of defects,  on a  surface of spherical topology,  decreases logarithmically with separation~\cite{Lubensky:1992p531} and  is  repulsive when their charges are of the same sign~\cite{Park:1992p946}.  We therefore expect, at  low temperatures,  to see four $+1/2$ disclinations at equal distance from each other.   In Fig.\ref{fig:geodesplot} this is evidenced  in  the  single peak  in the  measured distribution of geodesic distances ($P(\xi)$) between the defects.   Directional bending stiffness ( $\kappa_{\parallel}\ne 0$ ) alters this distribution considerably, since curvature influences the  interaction between defects~\cite{Bowick:2009p955,Vitelli:2006p1462}.  We see that  two of the $+1/2$ defects move towards each other to  form  pairs as the  vesicle deforms into a tube, resulting in a peak in the $\xi$ distribution at short separation as shown in Fig.\ref{fig:geodesplot}. This peak moves towards the left as $C_0^{\parallel}$ increases.  The pairs, which now has a total strength of $+1$, moves away from each other, resulting in a second peak in $P(\xi)$. This peak position moves to the right as $C_0^{\parallel}$ is increased.  We thus observe  nontrivial changes in the interaction between disclinations due to the geometry of the embedding surface. For $C_0^{\parallel}<0$, there is a proliferation of defects as the inward tubes are formed. The inner side of the tubes are decorated with ${\hat m}$ along the circumference   with the tip  hosting a pair of $+1/2$ defects. There is a sudden change in the orientation of ${\hat m}$ just out side the rim of the tube(see Fig.\ref{fig:line}). The rim itself is thus a line of discontinuity. There are two types of rims, closed ones with four  $-1/2$ defects  outside and paired open rims ending in two $-1/2$ defects.   

\begin{figure}
\centering
\includegraphics[width=5in]{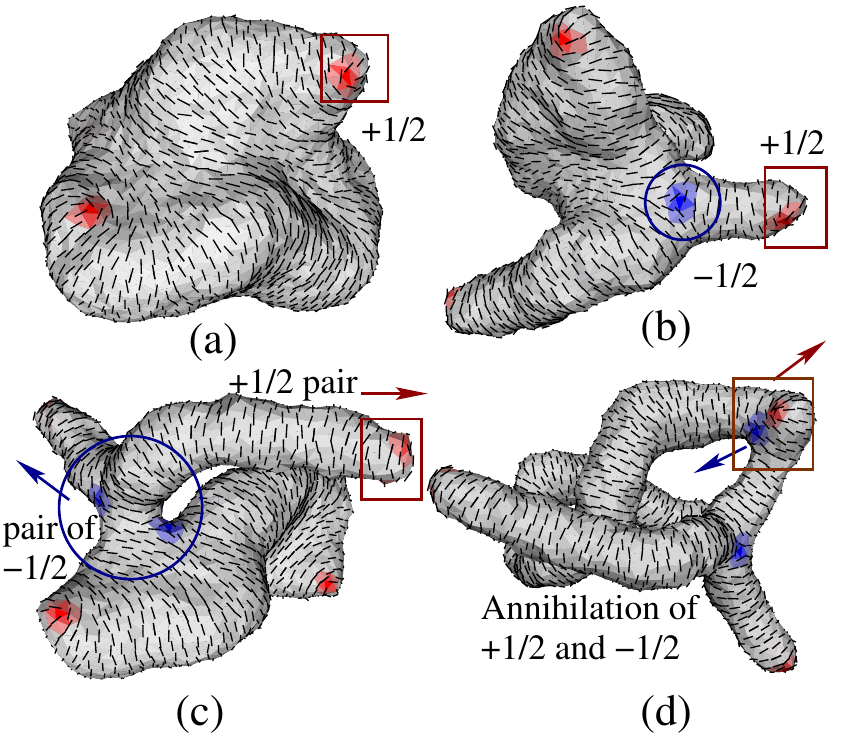}
\caption{\label{fig:spiral-kin} (Color online) The formation of tubes and branches are  driven by proliferation of a +1/2 and -1/2 defect pairs . See text for details.}
\end{figure}
   
A variety of thermally excited shapes are also seen.  The lowest energy thermal modes among the tubes  are the spirals,  which  are always present unless one is at a very low temperature (see figure 1f). If we raise  the temperature  or decrease the value of $\epsilon_{LL}$,  the tubes develop  branches.  More branches  are produced as the temperature is increased, with every branch having a pair of $-1/2$  defects at the intersection and a pair of   $+1/2$ defects at the tip~\cite{supp}. Fig.\ref{fig:spiral-kin} shows the sequence of snap shots that depict  the transformation of a sphere with an initial random orientation of the nematic to a tube,  when quenched from $\epsilon_{LL}=0$ to $\epsilon_{LL}=3$, while keeping $ \kappa_{\parallel}=5, \kappa_{\perp}=0.0, \kappa=10$ and $C_0^{\parallel}=0.4$~\cite{footnote}. The nematic instantly orients  on the sphere resulting in four $+ 1/2$ defects at equal distance from each other.  The membrane then begins to deform, through formation of protrusions at the defect cores (Fig.\ref{fig:spiral-kin}(a)), since positive defects favor positively curved regions\cite{Bowick:2009p955} and to satisfy the curvature set by $C_{0}^{\parallel}$ and $\kappa$.   These  membrane protrusions are accompanied by  production  of oppositely charged  half defect pairs(Fig.\ref{fig:spiral-kin}(b)).  The negative defects move to the negatively curved intersections and the positive defects move to the positively curved tip of the protrusions.   When the tip of a protrusion  has two $+1/2$ defects, and its neck two $-1/2$,  that branch is stabilized  and grows by moving the positive and negative defect pairs  away from each other(Fig.\ref{fig:spiral-kin}(c)). Isolated  $+1/2$ and $-1/2$ defects annihilate(Fig.\ref{fig:spiral-kin}(d)).  In general two types of branched structures are observed, as we increase the directional spontaneous curvature,  (a) {\it Broad necked}, see Fig.\ref{fig:kperp0}(g) and (b) {\it Narrow necked}, see Fig.\ref{fig:kperp0}(h) .
 
To analyze the origin of helicity, we neglect  the end cap effects of the tube  and parameterize it as  a {\it canal surface}.   The total  elastic energy, per unit length, is then~\cite{supp}
 {\small $ {\cal H}_{\rm tot} = \frac{\pi}{r} \left(\epsilon_{LL} \left(1-\sqrt{1-(\lambda r)^2}\right)+  
 \left(\frac{\kappa + \kappa_{\parallel}(1-x)^2 +\kappa_{\perp} x^2} {\sqrt{1 -\left(\lambda r\right)^2}}\right) 
  + \left(\frac{\kappa_{\parallel}}{2}(1-C_{0}^{ \parallel} r)(2x-1-HC_{0}^{\parallel} r) +  \frac{\kappa_{\perp}}{2}(1-C_{0}^{\perp}r)
 (1-2x-C_{0}^{\perp}r)\right) \right )$}.  Here $r$ is the radius of the tube and $x=\cos^2 \varphi$.  The energy depends only  on the curvature of its spine curve $\lambda$, and not on the torsion, hence for $\lambda \neq 0$ there is a degeneracy of spiral configurations corresponding to different torsions. One can easily see that the energy is at its minimum when $\lambda =0$~\cite{supp}. For a tube of length $L$, parametrizing the spiral deformations  by the extension of the tube along the helical axis $L_z=L\sqrt{1-\lambda r}$, we can estimate the entropy to be  of the  order $k_B T \ln(\frac{L}{a})$ and for small  $\lambda$ values the entropy term will dominate the free energy resulting in helical configurations. 

Before we conclude, let us look at how does the model parameters like the spontaneous curvature and the density of nematic inclusions compare with experiments.  For  $C_0^{\parallel} \sim 1.0$, the circumference of the tube is about 4 tether lengths.  Comparing this with the radius of the lipid tubes obtained in experiments \cite{Sweitzer:1998p3632} we get the length of the tether to be $\approx 25$nm. The value of $C_0^{\parallel}=1.0$ in real units is thus $\approx (25 ~\rm nm)^{-1}$, which is not far from the suggested value of dynamin intrinsic curvature.  Experiments  see the dynamin rings surrounding the tubes to be  made of about 20 units with  a pitch of about $15$ nm. This  will translate to about 5 dynamin molecules per vertex in the simulations. 

In conclusion, we have shown that in-plane nematic order couple to curvature, on a deformable surface, can lead to non-trivial shapes of vesicles.  The deformability of the surface lead to generation of point defects and line singularities,which in turn leads to production of tubes and branches.  


\begin{thebibliography}{10}

\bibitem{Markowitz:1991p3276}
M. Markowitz and A. Singh, Langmuir {\bf 7},  16  (1991).

\bibitem{Praefcke:2004p3309}
G.~J.~K. Praefcke and H.~T. McMahon, Nat. Rev. Mol. Cell Biol. {\bf 5},  133
  (2004).

\bibitem{Zimmerberg:2006p510}
J. Zimmerberg and M.~M. Kozlov, Nat. Rev. Mol. Cell Biol. {\bf 7},  9  (2006).

\bibitem{Voeltz:2007p1399}
G.~K. Voeltz and W.~A. Prinz, Nat. Rev. Mol. Cell Biol. {\bf 8},  258  (2007).

\bibitem{Shibata:2009p643}
Y. Shibata, J. Hu, M.~M. Kozlov, and T.~A. Rapoport, Annu. Rev. Cell Dev. Biol.
  {\bf 25},  329  (2009).

\bibitem{Helfrich:1973p693}
W. Helfrich, Z. Naturforsch. C {\bf 28},  693  (1973).

\bibitem{Fournier:1996p488}
J.~B. Fournier, Phys. Rev. Lett. {\bf 76},  4436  (1996).

\bibitem{Fournier:1998p2999}
J.-B. Fournier and P. Galatola, Brazilian J. Phys. {\bf 28},  329  (1998).

\bibitem{KraljIglic:2006p3974}
V. Kralj-Igli{\v c} {\it et~al.}, J. Stat. Phys. {\bf 125},  727  (2006).

\bibitem{Smith:1988p3535}
G.~S. Smith, E.~B. Sirota, C.~R. Safinya, and N.~A. Clark, Phys. Rev. Lett.
  {\bf 60},  813  (1988).

\bibitem{Tardieu:1973p711}
A. Tardieu, V. Luzzati, and F.~C. Reman, J. Mol. Bio. {\bf 75},  711   (1973).

\bibitem{Yin:2009p255}
Y. Yin, A. Arkhipov, and K. Schulten, Structure {\bf 17},  882  (2009).

\bibitem{Park:1992p946}
J. Park, T.~C. Lubensky, and F.~C. Mackintosh, Europhys. Lett. {\bf 20},  279
  (1992).

\bibitem{MacKintosh:1991p685}
F.~C. MacKintosh and T.~C. Lubensky, Phys. Rev. Lett. {\bf 67},  1169  (1991).

\bibitem{Helfrich:1988p3004}
W. Helfrich and J. Prost, Phys. Rev. A {\bf 38},  3065  (1988).

\bibitem{Nelson:1992p9}
P. Nelson and T. Powers, Phys. Rev. Lett. {\bf 69},  3409  (1992).

\bibitem{Selinger:1993p108}
J.~V. Selinger and J.~M. Schnur, Phys. Rev. Lett. {\bf 71},  4091  (1993).

\bibitem{Schnur:1993p1277}
J.~M. Schnur, Science {\bf 262},  1669  (1993).

\bibitem{Ayton:2007p3485}
G.~S. Ayton, P.~D. Blood, and G.~A. Voth, Biophysical Journal {\bf 92},  3595
  (2007).

\bibitem{Ayton:2009p3487}
G.~S. Ayton {\it et~al.}, Biophys. J. {\bf 97},  1616  (2009).

\bibitem{Shin:2008p174}
H. Shin, M. Bowick, and X. Xing, Phys. Rev. Lett. {\bf 101},  037802  (2008).

\bibitem{Nelson:2002p335}
D.~R. Nelson, Nano Lett. {\bf 2},  1125  (2002).

\bibitem{Ramakrishnan:2010p1249}
N. Ramakrishnan, P.~B.~S. Kumar, and J.~H. Ipsen, Phys. Rev. E {\bf 81},  41922
   (2010).

\bibitem{supp}
See supplementary materials at \url{http://www.rsc.org/suppdata/sm/c2/c2sm07384f/c2sm07384f.pdf}

\bibitem{Lebwohl:1972p426}
P.~A. Lebwohl and G. Lasher, Phys. Rev. A {\bf 6},  426  (1972).

\bibitem{Gompper:1994p205}
G. Gompper and D.~M. Kroll, Phys. Rev. Lett. {\bf 73},  2139  (1994).

\bibitem{Lubensky:1992p531}
T.~C. Lubensky and J. Prost, J. Phys. II France {\bf 2},  371  (1992).

\bibitem{Bowick:2009p955}
M. Bowick and L. Giomi, Adv. Phys. {\bf 58},  449  (2009).

\bibitem{Vitelli:2006p1462}
V. Vitelli and D.~R. Nelson, Phys. Rev. E {\bf 74},  21711  (2006).

\bibitem{footnote}  Such a quench can be achieved  by  a change in temperature across  the isotropic-nematic  transition, wherein the mean bending modulus remains almost the same but  a sharp change in the nematic stiffness.

\bibitem{Sweitzer:1998p3632}
S.~M. Sweitzer and J.~E. Hinshaw, Cell {\bf 93},  1021  (1998).

\end{thebibliography}
\bibliographystyle{prsty} 

\end{document}